\documentclass[aps,twocolumn,prx,superscriptaddress,nofootinbib]{revtex4-1}
\pdfoutput=1
\usepackage{amsmath}

\usepackage{amssymb}
\usepackage{mathrsfs}
\usepackage{bm, dsfont}
\usepackage[usenames,dvipsnames]{color}
\usepackage{colortbl}
\usepackage[table]{xcolor}
\usepackage{enumitem}

\usepackage{graphicx}
\usepackage{natbib}
\usepackage[colorlinks=true,linkcolor=blue,citecolor=blue,urlcolor=blue]{hyperref}
\usepackage{cleveref}
\usepackage{hypcap}
\usepackage{verbatim, float}
\usepackage{psfrag}
\usepackage[normalem]{ulem}
\usepackage{physics}		

\newcommand{\id}{\mathds{1}}
\newcommand{\ii}{\mathrm{i}}

\newcommand{\cE}{\mathcal{E}}

\newcommand{\cH}{\mathcal{H}}

\newcommand{\cL}{\mathcal{L}}
\newcommand{\cW}{\mathcal{W}}

\renewcommand{\t}[1]{\mathrm{#1}}
\newcommand{\be}{\begin{equation}}
\newcommand{\ee}{\end{equation}}

\newcommand{\nc}{\textup{\,\o\,}}

\begin{document}

\title{Unlimited One-Way Steering}

\author{Pavel Sekatski}
\email{pavel.sekatski@unige.ch}
\author{Florian Giraud}
\author{Roope Uola}
\author{Nicolas Brunner}
\affiliation{Department of Applied Physics, University of Geneva, Switzerland}

\begin{abstract}
    This work explores the asymmetry of quantum steering in a setup using high-dimensional entanglement. We construct entangled states with the following properties: $(i)$ one party (Alice) can never steer the state of the other party (Bob), considering the most general measurements, and $(ii)$ Bob can strongly steer the state of Alice, demonstrating genuine high-dimensional steering. In other words, Bob can convince Alice that they share an entangled state of arbitrarily high Schmidt number, while Alice can never convince Bob that the state is even simply entangled. In this sense, one-way steering can become unlimited. A key result for our construction is a condition for the joint measurability of all high-dimensional measurements subjected to the combined effect of noise and loss, which is of independent interest. 
\end{abstract}

\maketitle

\section{Introduction}

Nonlocality is among the central features of quantum theory. This effect manifests at different levels. On the one hand, in the mathematical structure of the theory (i.e. the Hilbert space) via the notion of entanglement \cite{HorodeckiReview,GuhneReview}. On the other hand, the measurement statistics of local measurements performed on entangled states can feature strong nonlocal correlations, which are incompatible with any local model \cite{BrunnerReview}. Initially, believed to be two facets of the same phenomenon, it is now clear that entanglement and quantum Bell nonlocality are in fact inherently different, see e.g. \cite{WernerState,Barrett2002}. 

Another perspective on quantum nonlocality is provided by the notion of quantum steering, see \cite{CavalcantiReview,UolaReview} for reviews. This effect, formalized by Wiseman, Jones and Doherty \cite{WisemanSteering}, takes its roots in the early works of Einstein-Podolsky-Rosen \cite{EPR} and Schr\"odinger \cite{schrodinger35,schrodinger36}. Here, an untrusted party (Bob) wants to convince another party (Alice) that they share an entangled state. By demonstrating his ability to remotely steer Alice’s local state in different measurement basis, Bob can convince Alice. Steering is thus an inherently asymmetrical task, contrary to entanglement and Bell nonlocality. 

Interestingly, the asymmetry of steering is also observed at the level of quantum states. Specifically, there exist entangled states that lead to steering from Bob to Alice, but not the other way around \cite{1W-EPR}; Bob can convince Alice that the shared state is entangled, while Alice can never convince Bob, even if she would use all possible local measurements. This effect, coined ``one-way steering’’ has attracted considerable attention in recent years, with many examples in low-dimensional (mostly two-qubit) \cite{1W-EPR,Paul1way,Quintino2015,Bowles1way2qubit,Zeng2022,Vertesi2021} and continuous variable Gaussian systems \cite{Midgley1wayGaussian,Olsen1wayGaussian} as well as experimental demonstrations, see e.g. \cite{Handchen2012,Wollmann2016,Sun2016,Tischler2018}.

A relevant question is whether there exists different forms of one-way steering, and whether some are stronger than others. So far, this question has not been discussed, due to the lack of an appropriate measure for steering in this context. Here, we tackle this problem, taking advantage of the recently introduced notion of genuine high-dimensional steering \cite{Designolle2021}, see also \cite{Designolle2022,degois2022complete}. This allows for a dimensional quantification of steering, specifically to lower bound the Schmidt number of an entangled state in a steering scenario. We then ask whether there exist entangled state with the following properties: (i) Alice cannot steer the state of Bob (even when allowing for all possible local measurements), and (ii) Bob can steer Alice's state strongly, i.e. ensuring the presence of high-dimensional entanglement (Schmidt number at least $d$). We answer this question in the affirmative by constructing a family of entangled states (of dimension $d\times (d+1) $) with the above two properties for any finite dimension $d$. 
To do so, we exploit the connection between steering and measurement incompatibility \cite{QuintinoSteerJM,UolaSteerJM1,UolaSteerJM2,KiukasSteerJM}, and also its recent generalisation \cite{jones2022} to high-dimensional steering and the concept of $n$-simulability of measurements \cite{ioannou2022}. Notably, we give a sufficient condition for the joint measurability of the set of all POVMs subjected to the combined effect of noise and loss.

\section{Question and main result}

We consider a scenario where two distant parties, Alice and Bob, share an entangled state $\varrho_{AB}$. Each party performs local measurements, represented by sets of 
positive-operator-valued measures (POVMs). Specifically, Alice's measurements are described by a set of POVMs $\{A_{a|x}\}$, where $x$ denotes the measurement choice and $a$ the outcome, with the properties that $A_{a|x}\geq0$ and $\sum_a A_{a|x}=\openone$ for all $a$ and $x$. Similarly, for Bob we define the set of POVMs $\{B_{b|y}\}$.

As the shared state is entangled, the effect of each party's local measurements is to remotely prepare (steer) the other party's state. This is described via two so-called state assemblages: $\{\sigma^A_{b|y}\}$ describe the states of Alice's system conditioned on Bob's measurement while $\{\sigma^B_{a|x}\}$ are Bob's states given Alice's measurement. These are given by
\begin{align} \label{assemblage}
\sigma^A_{b|y}=\text{tr}_B[(  \openone \otimes B_{b|y} )\varrho_{AB}] \,, 
\\
\sigma^B_{a|x}=\text{tr}_A[(   A_{a|x}  \otimes \openone )\varrho_{AB}] \,.
\end{align}

 The main question we address here is how different these two state assemblages can become, in other words how asymmetric the effect of steering can be. Loosely speaking we are looking for an entangled state $\varrho_{AB}$ such that one of the state assemblages, say $\{\sigma^A_{b|y}\}$, is classical (in the sense that it can never lead to steering), while the other assemblage $\{\sigma^B_{a|x}\}$ is highly non-classical (in the sense that it exhibits strong steering, witnessing high entanglement dimensionality).

 Before defining the problem more precisely, let us first observe that we are looking for some entangled states that are high-dimensional and asymmetrical. Obviously, if the state $\varrho_{AB}$ would be symmetrical under the exchange of Alice and Bob, any assemblage obtainable in one direction can also be obtained the other way around. Moreover, the state $\varrho_{AB}$ should feature a high entanglement dimensionality, as quantified here via the Schmidt number \cite{Terhal,Sanpera}: the Schmidt number of $\varrho_{AB}$ is the minimum $n$ such that there exists a decomposition $ \varrho_{AB} = \sum_j p_j \ket{\psi_j}\bra{\psi_j}$ where all $\ket{\psi_j} $ are pure entangled states of Schmidt rank at most $n$.
 
 More formally, we are looking for entangled states with the following two properties:
 
 \begin{enumerate}
     \item The assemblage $\{\sigma^A_{b|y}\}$ admits a local hidden state (LHS) model \cite{WisemanSteering}: 
     \be 
     \sigma^A_{b|y} =  p(b|y)\sum_\lambda p(\lambda|b,y)\sigma_\lambda \quad \forall b,y \,.
     \ee 
     Importantly, this should hold for any set of measurements for Bob $\{B_{b|y}\}$. Thus, no steering from Bob to Alice is possible, implying that Bob can never convince Alice that the shared state $\varrho_{AB}$ is entangled.
     \item The assemblage $\{\sigma^B_{a|x}\}$ is not $(d-1)$-preparable \cite{Designolle2021}, i.e. 
     \be
     \sigma^B_{a|x} \neq \sum_\lambda p(\lambda)\text{tr}_A[(   M^{\lambda}_{a|x}  \otimes \openone )\rho^{\lambda}_{AB}] 
     \ee
     with $p(\lambda)$ an arbitrary probability distribution, $M^{\lambda}_{a|x}$ arbitrary measurements (in $\mathbb{C}^d$), and all $\rho^{\lambda}_{AB}$ being arbitrary states of Schmidt number (at most) $d-1$. Thus, genuine $d$-dimensional steering is demonstrated, so that Alice can convince Bob that the shared state $\varrho_{AB}$ has large Schmidt number $d$. For short, we say that $\varrho_{AB}$ is $d$-steerable.
 \end{enumerate}

In the following, we construct a family of entangled states in dimension $d \times (d+1)$ with the above two properties, for any $d$. This shows that one-way steering can become unlimited, in the sense of being maximally asymmetrical. Specifically, we start from a maximally entangled two-qudit state $\ket{\phi^+_d}=\frac{1}{\sqrt{d}} \sum_{k=0}^{d-1} \ket{k,k}$ (with the notation $\Phi_d^+ = \ketbra{\phi^+_d}$) and apply successively a white noise (depolarizing) channel followed by a loss channel, defined by
\begin{align} \label{eq: white noise}
    \cW_p &:  \rho \mapsto p \,\rho + (1-p) \text{tr}[\rho]\frac{\id_d}{d} \,,
    \\
    \label{eq: loss}
    \cL_\eta &: \rho \mapsto \eta \,\rho + (1-\eta) \text{tr}[\rho]\ketbra{\nc}.
\end{align}
The depolarizing (loss) channel replaces the original state with a maximally mixed state (vacuum state $\ketbra{\nc}$) with probability $1-p$ ($1-\eta$). Note that the vacuum state represents an additional "level", orthogonal to the input Hilbert space, so that the output state has dimension $d+1$. After the two channels, we obtain the final state 
\begin{align}\label{eq: state}
\varrho_{AB}^{(\eta,p)} &= \t{id}\otimes (\cL_\eta\circ \cW_p)[\Phi^+] \\
    & = \eta p\, \Phi^+_d + \eta(1-p)\frac{\id_{d}\otimes \id_d}{d^2} + (1-\eta) \frac{\id_d}{d}\otimes\ketbra{\nc} \,,
    \nonumber
\end{align}
which is of dimension $d \times (d+1)$.\\

\noindent \textbf{Result 1.} {\it The states $\varrho_{AB}^{(\eta,p)}$ defined in Eq.~\eqref{eq: state} satisfy the following properties}\\

    {\it
    (i)  $\varrho_{AB}^{(\eta,p)}$ is $d$-steerable from Alice to Bob if 
    \be \label{eq: bound d-steer}
    p >  \frac{d\sqrt{\frac{d}{d+1}}-1}{d-1}.
    \ee
      
      (ii)  $\varrho_{AB}^{(\eta,p)}$ is unsteerable from Bob to Alice if 
      \be
      \eta \leq (1-p)^{d-1}.
      \ee

\noindent Therefore, for any $d$ there are parameter values $(\eta,p)$ such that conditions $(i)$ and $(ii)$ are both satisfied, thus demonstrating unlimited one-way steering. This regime corresponds to low noise and high loss.}

\section{Proof outline}

\subsection{Statement (i)}

 The first step is to show that the loss channel can be removed, because when acting on the trusted party it does not affect the steering properties of the underlying assemblage. This statement is formalized by the following lemma.\\

\noindent\textbf{Lemma 1.} \textit{Consider an assemblage $\{\sigma_{a|x}\}_{a,x}$ and a loss channel $\cL_\eta$ for $\eta>0$. The assemblage  $\left\{\sigma'_{a|x}=  \cL_\eta[\sigma_{a|x}] \right\}_{a,x} $ is $n$-steerable if and only if $\{\sigma_{a|x}\}_{a,x}$ is.}

 The full proof is in Appendix~\ref{app: lemma 1}. The idea is to show that if the assemblage $\left\{\sigma'_{a|x} \right\}_{a,x}$ is $n$-preparable, one can construct an $n$-preparable model for $\{\sigma_{a|x}\}_{a,x}$. The converse holds trivially (via application of the loss channel).

Hence, to prove that $\rho^{(\eta,p)}_{AB}$ is $d$-steerable from Bob to Alice it is sufficient to show that the isotropic state
$\rho^{(1,p)}_{AB} = p \, \Phi^+_d + (1-p) \frac{\id_{d}\otimes \id_d}{d^2}
$ is, which is to be expected for weak enough noise. Indeed, it was shown in \cite{jones2022} that this is the case for $p$ satisfying Eq. \eqref{eq: bound d-steer}, concluding the proof of statement $(i)$.
 Note that, in the above, $d$-steerability is demonstrated when using the set of all measurements. Instead, one could use a simpler (and practical) witness based on a pair of MUBs, certifying $d$-steerability of $\rho_{AB}^{(1,p)}$ \cite{Designolle2021} for 

\be\label{eq: bound 2 MUBs}
p\geq \frac{(d+\sqrt{d}-1)\sqrt{d-1}-1}{(d-1)(\sqrt{d-1}+1)}.
\ee
In turn, one shows $d$-steerability of $\rho_{AB}^{(\eta,p)}$ for any $\eta>0$, via the filtering procedure discussed in the proof of the lemma 2, which removes the vacuum component $\ket{\nc}$.

\subsection{Statement (ii)}

Here we exploit the connection~\cite{UolaSteerJM2} between the notions of steering and measurement compatibility, formally introduced in the next section. In particular, for any CPTP map $\cE$ it is known that a state assemblage given by the operators $\sigma_{b|y}= \tr[ (\id \otimes M_{b|y})( \t{id}\otimes \cE [\Phi^+])]$ admits a LHS model iff the set of measurements $\{\cE^*[M_{b|y}]\}$ is jointly measurable. Hence, the state 
$( \t{id}\otimes \cE [\Phi^+])$ is unsteerable if and only if the channel $\cE$ is  
\textit{incompatibility breaking}~\cite{HeinosaariIBC}, that is for any measurement assemblage $\{M_{b|y}\}$ the resulting assemblage $\{\cE^*[M_{b|y}]\}$ is jointly measurable. Thus, statement $(ii)$ is implied by the following lemma.\\

\noindent \textbf{Lemma 2} {\it The channel $\cE=\cL_\eta\circ \cW_p$ is incompatibility breaking if $\eta \leq (1-p)^{d-1}$.}\\

 The proof of the lemma is rather straightforward once we have solved the question of joint measurability of all measurements subject to white noise and loss in a given finite dimension $d$. This is a question of independent interest, and will be exposed in the next section. Then these results are used in Appendix \ref{app: lemma 2} to give a formal proof of Lemma 2. 

\section{Compatibility of all measurements with losses and noise}

Measurement incompatibility is a property of a set of measurements which cannot be performed simultaneously on a single copy of a system. Formally, a set of measurements $\{M_{a|x}\}$ is called incompatible if there exists no "parent" POVM $\{E_\lambda\}$ and classical post-processings $\{p(a|x,\lambda)\}$ such that
\begin{align}\label{eq: response funcion}
M_{a|x}=\sum_\lambda p(a|x,\lambda)E_\lambda.
\end{align}
Sets of measurements allowing for such a model are called jointly measurable; see e.g. \cite{HeinosaariIncompatibilityReview,GuhneJMColloquium} for reviews.

Our focus is on the (in)compatibility of measurements which are noisy and lossy. For any POVM $\{M_a\}$ with $m$ outputs $a=1,\dots,m$ acting on a system of dimension $d$, we define its imperfect version $\{\bar M_a^{(\eta ,p)}\}$ as a POVM with $m+1$ elements given by
\be
\bar M_a^{(\eta,p)} = 
\begin{cases}
 \eta p \, M_{a} + \eta (1-p) (\tr M_a) \frac{\id_d}{d} & a=1,\dots,m  \\ 
 (1-\eta) \id_d & a= \nc
\end{cases}.
\label{eq: noisy POVM}\ee
The noisy measurement apparatus behaves like the ideal one with probability $\eta p$,  produces a random outcome with probability $\eta(1-p)$, and does not click with probability $1-\eta$ (formally this corresponds to the no-click outcome $a=\nc$).  We are interested in knowing for which values of $(\eta,p)$ all measurements in a given dimension $d$ become compatible. The following result gives a sufficient condition.

\noindent \textbf{Result 2}. {\it The set of all POVMs $\{\bar M_{a|U}^{(\eta,p)}\}_U$ on $\mathds{C}^d$ is jointly measurable if
 \be\label{eq: JM condition}
 \eta \leq (1-p)^{d-1}.
 \ee
}

\textit{Proof:} To prove the result we present an explicit construction, inspired by~\cite{hirsch2013}, able to simulate all imperfect POVMs with noise parameters fulfilling Eq.~\eqref{eq: JM condition}. Note that it is sufficient to show that it can simulate all rank-one POVMs $\{M_a = \alpha_a \ketbra{\varphi_a}\}$ with $\sum \alpha_a = d$, since they can be post-processed to simulate all other measurements. The post-processing is done by mixing POVM elements, i.e. coarse-graining the corresponding outputs, by linearity it is consistent with the noisification of the measurements defined in Eq.~\eqref{eq: noisy POVM}.

We take the parent POVM to be the covariant one -- the continuous-valued measurement with the density
\be\label{eq main: parent covariant}
E_{\bm z} = d \ketbra{\bm z}
\ee
where $\ket{\bm z} =\sum_{k=0}^{d-1} z_k \ket{k}$ with $\bm z\in \mathds{C}^d$ and $|\bm z|^2=1$,    and $\dd \bm z$ is the invariant (under unitary transformations) measure over pure quantum states $\ket{\bm z}$ such that $\int \dd \bm z \ketbra{\bm z} = \frac{1}{d} \id_d$.


To define the response function $p(a|\bm z)$ simulating a notified rank-one POVM $\{\alpha_a \ketbra{\varphi_a}\}_a$ we proceed in two steps. First one samples a possible output $a$ from the probability distribution $ \frac{\alpha_a}{d}$. Given $a$ and the corresponding state $\ket{\varphi_a}$, one 
simulates a two outcome POVM $\{N_a^{(a)},N_\nc^{(a)}\}$ by using the deterministic response functions
\be
p^{(a)}(a|\bm z) = \begin{cases} 1 & |\braket{\varphi_a}{\bm z}|^2\geq t\\ 
0 & \text{otherwise}
\end{cases}
\ee
and 
$p^{(a)}(\nc|\bm z)= 1-p^{(a)}(a|\bm z)$, with a parameter $t\in[0,1]$.  Let us now compute the resulting operators 
\be
N_a^{(a)} = \int \dd \bm z\,  p^{(a)}(a|\bm z) E_{\bm z}
\ee
and $N_\nc^{(a)}= \id_d -N_a^{(a)}$. One notes that $N_a^{(a)} = U N_a^{(a)} U^\dag$ is invariant under  all unitary transformations $U$ which leave the state $\ket{\varphi_a}$ unchanged $U\ket{\varphi_a} =\ket{\varphi_a}$, since both the measure $\dd \bm z$ and the response function $p^{(a)}(a|\bm z)$ are invariant under such transformations. In other words, any such unitary commutes with our operator $[U,N_a^{(a)}]=0$. It follows that $N_a^{(a)}$ has $\ket{\varphi_a}$ for an eigenstate, and is furthermore proportional to identity on the orthogonal subspace. I.e. it is of the form
\be
N_a^{(a)} = \bar A_d(t) \ketbra{\varphi_a} + \bar B_d(t) \frac{\id_d-\ketbra{\varphi_a}}{d-1},
\ee
with the scalar functions that can be computed as 
\be\begin{split}
\bar A_d(t) = \tr   N_a^{(a)}  \ketbra{\varphi_a} &=  d \int \dd \bm z\,  p^{(a)}(a|\bm z) |\braket{\varphi_a}{\bm z}|^2 \\
\bar A_d(t) + \bar B_d(t)  = \tr N_a^{(a)} &= d \int \dd \bm z\,  p^{(a)}(a|\bm z).
\end{split}
\ee
These integrals are straightforward to compute. In the Appendix~\ref{app: integrals} we show that they give $\bar A_d(t)=(1-t)^{d-1}\left((d-1)t+1 \right)$ and $\bar T_d(t) \equiv\bar A_d(t) +\bar B_d(t)= d(1-t)^{d-1}$.

Finally, averaging over the sampled value $a$ we see that this strategy simulates $ N_{a}= \frac{\alpha_a}{d} N_{a}^{(a)}$  and $N_\nc = \sum_{a} \frac{\alpha_a}{d} N_\nc^{(a)}$, leading to 
\be\label{eq: final simul}
N_a = \frac{\alpha_a}{d}\left(\bar A_d(t) \ketbra{\varphi_a} + \bar B_d(t) \frac{\id_d-\ketbra{\varphi_a}}{d-1}\right)
\ee
and $N_\nc = \id_d -\sum_a N_a$. On the other hand from Eq.~\eqref{eq: noisy POVM} we obtain $\bar M_a^{(\eta,p)} =\frac{\alpha_d}{d}\big(d \, \eta p \ketbra{\varphi_a}+ \eta (1-p)\id_d \big)$.
Comparing with Eq.~\eqref{eq: final simul} we conclude that for
\be\begin{split}
\eta &= \frac{\bar T_d(t)}{d}= (1-t)^{d-1},\\
p &= \frac{d \bar A_d(t) - \bar T_d(t)}{(d-1) \bar T_d(t) } = t
\end{split}
\ee 
and $t\in[0,1]$ any POVM $\{\bar M_a^{(\eta,p)}\}$ can be simulated. Reproducing measurements that are even more noisy is straightforward by adding noise to the above construction. Hence, for any noise level $p=t$ our construction simulates all POVMs $\{\bar M_a^{(\eta,p)}\}$ if $\eta \leq (1-t)^{d-1}=(1-p)^{d-1}$. $\square$

\section{Discussion and conclusion}

Considering a bipartite steering scenario based on high-dimensional entanglement, we have investigated how asymmetrical the effect of steering can become. We presented entangled states $\varrho_{AB}$ such that Alice can convince Bob that $\varrho_{AB}$ is of high Schmidt number, while Bob can never convince Alice that the state is even entangled. Thus, one-way steering can become unlimited. 

Specifically, we constructed families of entangled states that exhibit genuine $d$-dimensional steering in one direction, while remaining unsteerable (under the most general measurements) in the other direction. We note that this construction can be straightforwardly generalized to states with genuine $n$-dimensional steering in one direction and unsteerable the other way around, for any $1<n\leq d$ \footnote{For this, one should simply adapt the bounds in Eqs \eqref{eq: bound d-steer} and \eqref{eq: bound 2 MUBs} to demand only $n-$steerability, following the results of \cite{jones2022} for \eqref{eq: bound d-steer} and \cite{Designolle2021} for \eqref{eq: bound 2 MUBs}.}. A practical implementation of such states should be feasible, e.g. with the setups of Refs \cite{Zeng2018,Srivastav2022} for an experimental demonstration of unlimited one-way steering.

Finally, our work also contributed to the characterisation of joint measurability in high-dimensional measurements. In particular, we obtained a criterion for the compatibility of arbitrary measurements subjected to both noise and loss. This result is of independent interest and may find other applications.

\emph{Acknowledgments.---} We acknowledge financial support from the Swiss National Science Foundation (projects 192244, Ambizione PZ00P2-202179, and NCCR SwissMAP). 

\bibliographystyle{apsrev4-1}

\bibliography{HDsteering}{}

\begin{thebibliography}{41}%
\makeatletter
\providecommand \@ifxundefined [1]{%
 \@ifx{#1\undefined}
}%
\providecommand \@ifnum [1]{%
 \ifnum #1\expandafter \@firstoftwo
 \else \expandafter \@secondoftwo
 \fi
}%
\providecommand \@ifx [1]{%
 \ifx #1\expandafter \@firstoftwo
 \else \expandafter \@secondoftwo
 \fi
}%
\providecommand \natexlab [1]{#1}%
\providecommand \enquote  [1]{``#1''}%
\providecommand \bibnamefont  [1]{#1}%
\providecommand \bibfnamefont [1]{#1}%
\providecommand \citenamefont [1]{#1}%
\providecommand \href@noop [0]{\@secondoftwo}%
\providecommand \href [0]{\begingroup \@sanitize@url \@href}%
\providecommand \@href[1]{\@@startlink{#1}\@@href}%
\providecommand \@@href[1]{\endgroup#1\@@endlink}%
\providecommand \@sanitize@url [0]{\catcode `\\12\catcode `\$12\catcode
  `\&12\catcode `\#12\catcode `\^12\catcode `\_12\catcode `\%12\relax}%
\providecommand \@@startlink[1]{}%
\providecommand \@@endlink[0]{}%
\providecommand \url  [0]{\begingroup\@sanitize@url \@url }%
\providecommand \@url [1]{\endgroup\@href {#1}{\urlprefix }}%
\providecommand \urlprefix  [0]{URL }%
\providecommand \Eprint [0]{\href }%
\providecommand \doibase [0]{http://dx.doi.org/}%
\providecommand \selectlanguage [0]{\@gobble}%
\providecommand \bibinfo  [0]{\@secondoftwo}%
\providecommand \bibfield  [0]{\@secondoftwo}%
\providecommand \translation [1]{[#1]}%
\providecommand \BibitemOpen [0]{}%
\providecommand \bibitemStop [0]{}%
\providecommand \bibitemNoStop [0]{.\EOS\space}%
\providecommand \EOS [0]{\spacefactor3000\relax}%
\providecommand \BibitemShut  [1]{\csname bibitem#1\endcsname}%
\let\auto@bib@innerbib\@empty
\bibitem [{\citenamefont {Horodecki}\ \emph {et~al.}(2009)\citenamefont
  {Horodecki}, \citenamefont {Horodecki}, \citenamefont {Horodecki},\ and\
  \citenamefont {Horodecki}}]{HorodeckiReview}%
  \BibitemOpen
  \bibfield  {author} {\bibinfo {author} {\bibfnamefont {R.}~\bibnamefont
  {Horodecki}}, \bibinfo {author} {\bibfnamefont {P.}~\bibnamefont
  {Horodecki}}, \bibinfo {author} {\bibfnamefont {M.}~\bibnamefont
  {Horodecki}}, \ and\ \bibinfo {author} {\bibfnamefont {K.}~\bibnamefont
  {Horodecki}},\ }\href {\doibase 10.1103/RevModPhys.81.865} {\bibfield
  {journal} {\bibinfo  {journal} {Rev. Mod. Phys.}\ }\textbf {\bibinfo {volume}
  {81}},\ \bibinfo {pages} {865} (\bibinfo {year} {2009})}\BibitemShut
  {NoStop}%
\bibitem [{\citenamefont {Gühne}\ and\ \citenamefont
  {T{\'{o}}th}(2009)}]{GuhneReview}%
  \BibitemOpen
  \bibfield  {author} {\bibinfo {author} {\bibfnamefont {O.}~\bibnamefont
  {Gühne}}\ and\ \bibinfo {author} {\bibfnamefont {G.}~\bibnamefont
  {T{\'{o}}th}},\ }\href {\doibase 10.1016/j.physrep.2009.02.004} {\bibfield
  {journal} {\bibinfo  {journal} {Physics Reports}\ }\textbf {\bibinfo {volume}
  {474}},\ \bibinfo {pages} {1} (\bibinfo {year} {2009})}\BibitemShut {NoStop}%
\bibitem [{\citenamefont {Brunner}\ \emph {et~al.}(2014)\citenamefont
  {Brunner}, \citenamefont {Cavalcanti}, \citenamefont {Pironio}, \citenamefont
  {Scarani},\ and\ \citenamefont {Wehner}}]{BrunnerReview}%
  \BibitemOpen
  \bibfield  {author} {\bibinfo {author} {\bibfnamefont {N.}~\bibnamefont
  {Brunner}}, \bibinfo {author} {\bibfnamefont {D.}~\bibnamefont {Cavalcanti}},
  \bibinfo {author} {\bibfnamefont {S.}~\bibnamefont {Pironio}}, \bibinfo
  {author} {\bibfnamefont {V.}~\bibnamefont {Scarani}}, \ and\ \bibinfo
  {author} {\bibfnamefont {S.}~\bibnamefont {Wehner}},\ }\href {\doibase
  10.1103/revmodphys.86.419} {\bibfield  {journal} {\bibinfo  {journal}
  {Reviews of Modern Physics}\ }\textbf {\bibinfo {volume} {86}},\ \bibinfo
  {pages} {419} (\bibinfo {year} {2014})}\BibitemShut {NoStop}%
\bibitem [{\citenamefont {Werner}(1989)}]{WernerState}%
  \BibitemOpen
  \bibfield  {author} {\bibinfo {author} {\bibfnamefont {R.~F.}\ \bibnamefont
  {Werner}},\ }\href {\doibase 10.1103/PhysRevA.40.4277} {\bibfield  {journal}
  {\bibinfo  {journal} {Phys. Rev. A}\ }\textbf {\bibinfo {volume} {40}},\
  \bibinfo {pages} {4277} (\bibinfo {year} {1989})}\BibitemShut {NoStop}%
\bibitem [{\citenamefont {Barrett}(2002)}]{Barrett2002}%
  \BibitemOpen
  \bibfield  {author} {\bibinfo {author} {\bibfnamefont {J.}~\bibnamefont
  {Barrett}},\ }\href {\doibase 10.1103/PhysRevA.65.042302} {\bibfield
  {journal} {\bibinfo  {journal} {Phys. Rev. A}\ }\textbf {\bibinfo {volume}
  {65}},\ \bibinfo {pages} {042302} (\bibinfo {year} {2002})}\BibitemShut
  {NoStop}%
\bibitem [{\citenamefont {Cavalcanti}\ and\ \citenamefont
  {Skrzypczyk}(2016)}]{CavalcantiReview}%
  \BibitemOpen
  \bibfield  {author} {\bibinfo {author} {\bibfnamefont {D.}~\bibnamefont
  {Cavalcanti}}\ and\ \bibinfo {author} {\bibfnamefont {P.}~\bibnamefont
  {Skrzypczyk}},\ }\href {\doibase 10.1088/1361-6633/80/2/024001} {\bibfield
  {journal} {\bibinfo  {journal} {Reports on Progress in Physics}\ }\textbf
  {\bibinfo {volume} {80}},\ \bibinfo {pages} {024001} (\bibinfo {year}
  {2016})}\BibitemShut {NoStop}%
\bibitem [{\citenamefont {Uola}\ \emph {et~al.}(2020)\citenamefont {Uola},
  \citenamefont {Costa}, \citenamefont {Nguyen},\ and\ \citenamefont
  {Gühne}}]{UolaReview}%
  \BibitemOpen
  \bibfield  {author} {\bibinfo {author} {\bibfnamefont {R.}~\bibnamefont
  {Uola}}, \bibinfo {author} {\bibfnamefont {A.~C.~S.}\ \bibnamefont {Costa}},
  \bibinfo {author} {\bibfnamefont {H.~C.}\ \bibnamefont {Nguyen}}, \ and\
  \bibinfo {author} {\bibfnamefont {O.}~\bibnamefont {Gühne}},\ }\href
  {\doibase 10.1103/revmodphys.92.015001} {\bibfield  {journal} {\bibinfo
  {journal} {Reviews of Modern Physics}\ }\textbf {\bibinfo {volume} {92}}
  (\bibinfo {year} {2020}),\ 10.1103/revmodphys.92.015001}\BibitemShut
  {NoStop}%
\bibitem [{\citenamefont {Wiseman}\ \emph {et~al.}(2007)\citenamefont
  {Wiseman}, \citenamefont {Jones},\ and\ \citenamefont
  {Doherty}}]{WisemanSteering}%
  \BibitemOpen
  \bibfield  {author} {\bibinfo {author} {\bibfnamefont {H.~M.}\ \bibnamefont
  {Wiseman}}, \bibinfo {author} {\bibfnamefont {S.~J.}\ \bibnamefont {Jones}},
  \ and\ \bibinfo {author} {\bibfnamefont {A.~C.}\ \bibnamefont {Doherty}},\
  }\href {\doibase 10.1103/physrevlett.98.140402} {\bibfield  {journal}
  {\bibinfo  {journal} {Physical Review Letters}\ }\textbf {\bibinfo {volume}
  {98}} (\bibinfo {year} {2007}),\ 10.1103/physrevlett.98.140402}\BibitemShut
  {NoStop}%
\bibitem [{\citenamefont {Einstein}\ \emph {et~al.}(1935)\citenamefont
  {Einstein}, \citenamefont {Podolsky},\ and\ \citenamefont {Rosen}}]{EPR}%
  \BibitemOpen
  \bibfield  {author} {\bibinfo {author} {\bibfnamefont {A.}~\bibnamefont
  {Einstein}}, \bibinfo {author} {\bibfnamefont {B.}~\bibnamefont {Podolsky}},
  \ and\ \bibinfo {author} {\bibfnamefont {N.}~\bibnamefont {Rosen}},\ }\href
  {\doibase 10.1103/PhysRev.47.777} {\bibfield  {journal} {\bibinfo  {journal}
  {Phys. Rev.}\ }\textbf {\bibinfo {volume} {47}},\ \bibinfo {pages} {777}
  (\bibinfo {year} {1935})}\BibitemShut {NoStop}%
\bibitem [{\citenamefont {Schrödinger}(1935)}]{schrodinger35}%
  \BibitemOpen
  \bibfield  {author} {\bibinfo {author} {\bibfnamefont {E.}~\bibnamefont
  {Schrödinger}},\ }\href {\doibase 10.1017/S0305004100013554} {\bibfield
  {journal} {\bibinfo  {journal} {Mathematical Proceedings of the Cambridge
  Philosophical Society}\ }\textbf {\bibinfo {volume} {31}},\ \bibinfo {pages}
  {555–563} (\bibinfo {year} {1935})}\BibitemShut {NoStop}%
\bibitem [{\citenamefont {Schrödinger}(1936)}]{schrodinger36}%
  \BibitemOpen
  \bibfield  {author} {\bibinfo {author} {\bibfnamefont {E.}~\bibnamefont
  {Schrödinger}},\ }\href {\doibase 10.1017/S0305004100019137} {\bibfield
  {journal} {\bibinfo  {journal} {Mathematical Proceedings of the Cambridge
  Philosophical Society}\ }\textbf {\bibinfo {volume} {32}},\ \bibinfo {pages}
  {446–452} (\bibinfo {year} {1936})}\BibitemShut {NoStop}%
\bibitem [{\citenamefont {Bowles}\ \emph {et~al.}(2014)\citenamefont {Bowles},
  \citenamefont {V\'ertesi}, \citenamefont {Quintino},\ and\ \citenamefont
  {Brunner}}]{1W-EPR}%
  \BibitemOpen
  \bibfield  {author} {\bibinfo {author} {\bibfnamefont {J.}~\bibnamefont
  {Bowles}}, \bibinfo {author} {\bibfnamefont {T.}~\bibnamefont {V\'ertesi}},
  \bibinfo {author} {\bibfnamefont {M.~T.}\ \bibnamefont {Quintino}}, \ and\
  \bibinfo {author} {\bibfnamefont {N.}~\bibnamefont {Brunner}},\ }\href
  {\doibase 10.1103/PhysRevLett.112.200402} {\bibfield  {journal} {\bibinfo
  {journal} {Phys. Rev. Lett.}\ }\textbf {\bibinfo {volume} {112}},\ \bibinfo
  {pages} {200402} (\bibinfo {year} {2014})}\BibitemShut {NoStop}%
\bibitem [{\citenamefont {Skrzypczyk}\ \emph {et~al.}(2014)\citenamefont
  {Skrzypczyk}, \citenamefont {Navascu\'es},\ and\ \citenamefont
  {Cavalcanti}}]{Paul1way}%
  \BibitemOpen
  \bibfield  {author} {\bibinfo {author} {\bibfnamefont {P.}~\bibnamefont
  {Skrzypczyk}}, \bibinfo {author} {\bibfnamefont {M.}~\bibnamefont
  {Navascu\'es}}, \ and\ \bibinfo {author} {\bibfnamefont {D.}~\bibnamefont
  {Cavalcanti}},\ }\href {\doibase 10.1103/PhysRevLett.112.180404} {\bibfield
  {journal} {\bibinfo  {journal} {Phys. Rev. Lett.}\ }\textbf {\bibinfo
  {volume} {112}},\ \bibinfo {pages} {180404} (\bibinfo {year}
  {2014})}\BibitemShut {NoStop}%
\bibitem [{\citenamefont {Quintino}\ \emph {et~al.}(2015)\citenamefont
  {Quintino}, \citenamefont {V\'ertesi}, \citenamefont {Cavalcanti},
  \citenamefont {Augusiak}, \citenamefont {Demianowicz}, \citenamefont
  {Ac\'{\i}n},\ and\ \citenamefont {Brunner}}]{Quintino2015}%
  \BibitemOpen
  \bibfield  {author} {\bibinfo {author} {\bibfnamefont {M.~T.}\ \bibnamefont
  {Quintino}}, \bibinfo {author} {\bibfnamefont {T.}~\bibnamefont {V\'ertesi}},
  \bibinfo {author} {\bibfnamefont {D.}~\bibnamefont {Cavalcanti}}, \bibinfo
  {author} {\bibfnamefont {R.}~\bibnamefont {Augusiak}}, \bibinfo {author}
  {\bibfnamefont {M.}~\bibnamefont {Demianowicz}}, \bibinfo {author}
  {\bibfnamefont {A.}~\bibnamefont {Ac\'{\i}n}}, \ and\ \bibinfo {author}
  {\bibfnamefont {N.}~\bibnamefont {Brunner}},\ }\href {\doibase
  10.1103/PhysRevA.92.032107} {\bibfield  {journal} {\bibinfo  {journal} {Phys.
  Rev. A}\ }\textbf {\bibinfo {volume} {92}},\ \bibinfo {pages} {032107}
  (\bibinfo {year} {2015})}\BibitemShut {NoStop}%
\bibitem [{\citenamefont {Bowles}\ \emph {et~al.}(2016)\citenamefont {Bowles},
  \citenamefont {Hirsch}, \citenamefont {Quintino},\ and\ \citenamefont
  {Brunner}}]{Bowles1way2qubit}%
  \BibitemOpen
  \bibfield  {author} {\bibinfo {author} {\bibfnamefont {J.}~\bibnamefont
  {Bowles}}, \bibinfo {author} {\bibfnamefont {F.}~\bibnamefont {Hirsch}},
  \bibinfo {author} {\bibfnamefont {M.~T.}\ \bibnamefont {Quintino}}, \ and\
  \bibinfo {author} {\bibfnamefont {N.}~\bibnamefont {Brunner}},\ }\href
  {\doibase 10.1103/PhysRevA.93.022121} {\bibfield  {journal} {\bibinfo
  {journal} {Phys. Rev. A}\ }\textbf {\bibinfo {volume} {93}},\ \bibinfo
  {pages} {022121} (\bibinfo {year} {2016})}\BibitemShut {NoStop}%
\bibitem [{\citenamefont {Zeng}(2022)}]{Zeng2022}%
  \BibitemOpen
  \bibfield  {author} {\bibinfo {author} {\bibfnamefont {Q.}~\bibnamefont
  {Zeng}},\ }\href {\doibase 10.1103/PhysRevA.106.032202} {\bibfield  {journal}
  {\bibinfo  {journal} {Phys. Rev. A}\ }\textbf {\bibinfo {volume} {106}},\
  \bibinfo {pages} {032202} (\bibinfo {year} {2022})}\BibitemShut {NoStop}%
\bibitem [{\citenamefont {M\'arton}\ \emph {et~al.}(2021)\citenamefont
  {M\'arton}, \citenamefont {Nagy}, \citenamefont {Bene},\ and\ \citenamefont
  {V\'ertesi}}]{Vertesi2021}%
  \BibitemOpen
  \bibfield  {author} {\bibinfo {author} {\bibfnamefont {I.}~\bibnamefont
  {M\'arton}}, \bibinfo {author} {\bibfnamefont {S.}~\bibnamefont {Nagy}},
  \bibinfo {author} {\bibfnamefont {E.}~\bibnamefont {Bene}}, \ and\ \bibinfo
  {author} {\bibfnamefont {T.}~\bibnamefont {V\'ertesi}},\ }\href {\doibase
  10.1103/PhysRevResearch.3.043100} {\bibfield  {journal} {\bibinfo  {journal}
  {Phys. Rev. Res.}\ }\textbf {\bibinfo {volume} {3}},\ \bibinfo {pages}
  {043100} (\bibinfo {year} {2021})}\BibitemShut {NoStop}%
\bibitem [{\citenamefont {Midgley}\ \emph {et~al.}(2010)\citenamefont
  {Midgley}, \citenamefont {Ferris},\ and\ \citenamefont
  {Olsen}}]{Midgley1wayGaussian}%
  \BibitemOpen
  \bibfield  {author} {\bibinfo {author} {\bibfnamefont {S.~L.~W.}\
  \bibnamefont {Midgley}}, \bibinfo {author} {\bibfnamefont {A.~J.}\
  \bibnamefont {Ferris}}, \ and\ \bibinfo {author} {\bibfnamefont {M.~K.}\
  \bibnamefont {Olsen}},\ }\href {\doibase 10.1103/PhysRevA.81.022101}
  {\bibfield  {journal} {\bibinfo  {journal} {Phys. Rev. A}\ }\textbf {\bibinfo
  {volume} {81}},\ \bibinfo {pages} {022101} (\bibinfo {year}
  {2010})}\BibitemShut {NoStop}%
\bibitem [{\citenamefont {Olsen}(2013)}]{Olsen1wayGaussian}%
  \BibitemOpen
  \bibfield  {author} {\bibinfo {author} {\bibfnamefont {M.~K.}\ \bibnamefont
  {Olsen}},\ }\href {\doibase 10.1103/PhysRevA.88.051802} {\bibfield  {journal}
  {\bibinfo  {journal} {Phys. Rev. A}\ }\textbf {\bibinfo {volume} {88}},\
  \bibinfo {pages} {051802} (\bibinfo {year} {2013})}\BibitemShut {NoStop}%
\bibitem [{\citenamefont {H{\"a}ndchen}\ \emph {et~al.}(2012)\citenamefont
  {H{\"a}ndchen}, \citenamefont {Eberle}, \citenamefont {Steinlechner},
  \citenamefont {Samblowski}, \citenamefont {Franz}, \citenamefont {Werner},\
  and\ \citenamefont {Schnabel}}]{Handchen2012}%
  \BibitemOpen
  \bibfield  {author} {\bibinfo {author} {\bibfnamefont {V.}~\bibnamefont
  {H{\"a}ndchen}}, \bibinfo {author} {\bibfnamefont {T.}~\bibnamefont
  {Eberle}}, \bibinfo {author} {\bibfnamefont {S.}~\bibnamefont
  {Steinlechner}}, \bibinfo {author} {\bibfnamefont {A.}~\bibnamefont
  {Samblowski}}, \bibinfo {author} {\bibfnamefont {T.}~\bibnamefont {Franz}},
  \bibinfo {author} {\bibfnamefont {R.~F.}\ \bibnamefont {Werner}}, \ and\
  \bibinfo {author} {\bibfnamefont {R.}~\bibnamefont {Schnabel}},\ }\href@noop
  {} {\bibfield  {journal} {\bibinfo  {journal} {Nature Photonics}\ }\textbf
  {\bibinfo {volume} {6}},\ \bibinfo {pages} {596} (\bibinfo {year}
  {2012})}\BibitemShut {NoStop}%
\bibitem [{\citenamefont {Wollmann}\ \emph {et~al.}(2016)\citenamefont
  {Wollmann}, \citenamefont {Walk}, \citenamefont {Bennet}, \citenamefont
  {Wiseman},\ and\ \citenamefont {Pryde}}]{Wollmann2016}%
  \BibitemOpen
  \bibfield  {author} {\bibinfo {author} {\bibfnamefont {S.}~\bibnamefont
  {Wollmann}}, \bibinfo {author} {\bibfnamefont {N.}~\bibnamefont {Walk}},
  \bibinfo {author} {\bibfnamefont {A.~J.}\ \bibnamefont {Bennet}}, \bibinfo
  {author} {\bibfnamefont {H.~M.}\ \bibnamefont {Wiseman}}, \ and\ \bibinfo
  {author} {\bibfnamefont {G.~J.}\ \bibnamefont {Pryde}},\ }\href {\doibase
  10.1103/PhysRevLett.116.160403} {\bibfield  {journal} {\bibinfo  {journal}
  {Phys. Rev. Lett.}\ }\textbf {\bibinfo {volume} {116}},\ \bibinfo {pages}
  {160403} (\bibinfo {year} {2016})}\BibitemShut {NoStop}%
\bibitem [{\citenamefont {Sun}\ \emph {et~al.}(2016)\citenamefont {Sun},
  \citenamefont {Ye}, \citenamefont {Xu}, \citenamefont {Xu}, \citenamefont
  {Tang}, \citenamefont {Wu}, \citenamefont {Chen}, \citenamefont {Li},\ and\
  \citenamefont {Guo}}]{Sun2016}%
  \BibitemOpen
  \bibfield  {author} {\bibinfo {author} {\bibfnamefont {K.}~\bibnamefont
  {Sun}}, \bibinfo {author} {\bibfnamefont {X.-J.}\ \bibnamefont {Ye}},
  \bibinfo {author} {\bibfnamefont {J.-S.}\ \bibnamefont {Xu}}, \bibinfo
  {author} {\bibfnamefont {X.-Y.}\ \bibnamefont {Xu}}, \bibinfo {author}
  {\bibfnamefont {J.-S.}\ \bibnamefont {Tang}}, \bibinfo {author}
  {\bibfnamefont {Y.-C.}\ \bibnamefont {Wu}}, \bibinfo {author} {\bibfnamefont
  {J.-L.}\ \bibnamefont {Chen}}, \bibinfo {author} {\bibfnamefont {C.-F.}\
  \bibnamefont {Li}}, \ and\ \bibinfo {author} {\bibfnamefont {G.-C.}\
  \bibnamefont {Guo}},\ }\href {\doibase 10.1103/PhysRevLett.116.160404}
  {\bibfield  {journal} {\bibinfo  {journal} {Phys. Rev. Lett.}\ }\textbf
  {\bibinfo {volume} {116}},\ \bibinfo {pages} {160404} (\bibinfo {year}
  {2016})}\BibitemShut {NoStop}%
\bibitem [{\citenamefont {Tischler}\ \emph {et~al.}(2018)\citenamefont
  {Tischler}, \citenamefont {Ghafari}, \citenamefont {Baker}, \citenamefont
  {Slussarenko}, \citenamefont {Patel}, \citenamefont {Weston}, \citenamefont
  {Wollmann}, \citenamefont {Shalm}, \citenamefont {Verma}, \citenamefont
  {Nam}, \citenamefont {Nguyen}, \citenamefont {Wiseman},\ and\ \citenamefont
  {Pryde}}]{Tischler2018}%
  \BibitemOpen
  \bibfield  {author} {\bibinfo {author} {\bibfnamefont {N.}~\bibnamefont
  {Tischler}}, \bibinfo {author} {\bibfnamefont {F.}~\bibnamefont {Ghafari}},
  \bibinfo {author} {\bibfnamefont {T.~J.}\ \bibnamefont {Baker}}, \bibinfo
  {author} {\bibfnamefont {S.}~\bibnamefont {Slussarenko}}, \bibinfo {author}
  {\bibfnamefont {R.~B.}\ \bibnamefont {Patel}}, \bibinfo {author}
  {\bibfnamefont {M.~M.}\ \bibnamefont {Weston}}, \bibinfo {author}
  {\bibfnamefont {S.}~\bibnamefont {Wollmann}}, \bibinfo {author}
  {\bibfnamefont {L.~K.}\ \bibnamefont {Shalm}}, \bibinfo {author}
  {\bibfnamefont {V.~B.}\ \bibnamefont {Verma}}, \bibinfo {author}
  {\bibfnamefont {S.~W.}\ \bibnamefont {Nam}}, \bibinfo {author} {\bibfnamefont
  {H.~C.}\ \bibnamefont {Nguyen}}, \bibinfo {author} {\bibfnamefont {H.~M.}\
  \bibnamefont {Wiseman}}, \ and\ \bibinfo {author} {\bibfnamefont {G.~J.}\
  \bibnamefont {Pryde}},\ }\href {\doibase 10.1103/PhysRevLett.121.100401}
  {\bibfield  {journal} {\bibinfo  {journal} {Phys. Rev. Lett.}\ }\textbf
  {\bibinfo {volume} {121}},\ \bibinfo {pages} {100401} (\bibinfo {year}
  {2018})}\BibitemShut {NoStop}%
\bibitem [{\citenamefont {Designolle}\ \emph {et~al.}(2021)\citenamefont
  {Designolle}, \citenamefont {Srivastav}, \citenamefont {Uola}, \citenamefont
  {Valencia}, \citenamefont {McCutcheon}, \citenamefont {Malik},\ and\
  \citenamefont {Brunner}}]{Designolle2021}%
  \BibitemOpen
  \bibfield  {author} {\bibinfo {author} {\bibfnamefont {S.}~\bibnamefont
  {Designolle}}, \bibinfo {author} {\bibfnamefont {V.}~\bibnamefont
  {Srivastav}}, \bibinfo {author} {\bibfnamefont {R.}~\bibnamefont {Uola}},
  \bibinfo {author} {\bibfnamefont {N.~H.}\ \bibnamefont {Valencia}}, \bibinfo
  {author} {\bibfnamefont {W.}~\bibnamefont {McCutcheon}}, \bibinfo {author}
  {\bibfnamefont {M.}~\bibnamefont {Malik}}, \ and\ \bibinfo {author}
  {\bibfnamefont {N.}~\bibnamefont {Brunner}},\ }\href {\doibase
  10.1103/PhysRevLett.126.200404} {\bibfield  {journal} {\bibinfo  {journal}
  {Phys. Rev. Lett.}\ }\textbf {\bibinfo {volume} {126}},\ \bibinfo {pages}
  {200404} (\bibinfo {year} {2021})}\BibitemShut {NoStop}%
\bibitem [{\citenamefont {Designolle}(2022)}]{Designolle2022}%
  \BibitemOpen
  \bibfield  {author} {\bibinfo {author} {\bibfnamefont {S.}~\bibnamefont
  {Designolle}},\ }\href {\doibase 10.1103/PhysRevA.105.032430} {\bibfield
  {journal} {\bibinfo  {journal} {Phys. Rev. A}\ }\textbf {\bibinfo {volume}
  {105}},\ \bibinfo {pages} {032430} (\bibinfo {year} {2022})}\BibitemShut
  {NoStop}%
\bibitem [{\citenamefont {de~Gois}\ \emph {et~al.}(2022)\citenamefont
  {de~Gois}, \citenamefont {Plávala}, \citenamefont {Schwonnek},\ and\
  \citenamefont {Gühne}}]{degois2022complete}%
  \BibitemOpen
  \bibfield  {author} {\bibinfo {author} {\bibfnamefont {C.}~\bibnamefont
  {de~Gois}}, \bibinfo {author} {\bibfnamefont {M.}~\bibnamefont {Plávala}},
  \bibinfo {author} {\bibfnamefont {R.}~\bibnamefont {Schwonnek}}, \ and\
  \bibinfo {author} {\bibfnamefont {O.}~\bibnamefont {Gühne}},\ }\href@noop {}
  {\enquote {\bibinfo {title} {Complete hierarchy for high-dimensional steering
  certification},}\ } (\bibinfo {year} {2022}),\ \Eprint
  {http://arxiv.org/abs/2212.12544} {arXiv:2212.12544 [quant-ph]} \BibitemShut
  {NoStop}%
\bibitem [{\citenamefont {Quintino}\ \emph {et~al.}(2014)\citenamefont
  {Quintino}, \citenamefont {V{\'{e}}rtesi},\ and\ \citenamefont
  {Brunner}}]{QuintinoSteerJM}%
  \BibitemOpen
  \bibfield  {author} {\bibinfo {author} {\bibfnamefont {M.~T.}\ \bibnamefont
  {Quintino}}, \bibinfo {author} {\bibfnamefont {T.}~\bibnamefont
  {V{\'{e}}rtesi}}, \ and\ \bibinfo {author} {\bibfnamefont {N.}~\bibnamefont
  {Brunner}},\ }\href {\doibase 10.1103/physrevlett.113.160402} {\bibfield
  {journal} {\bibinfo  {journal} {Physical Review Letters}\ }\textbf {\bibinfo
  {volume} {113}} (\bibinfo {year} {2014}),\
  10.1103/physrevlett.113.160402}\BibitemShut {NoStop}%
\bibitem [{\citenamefont {Uola}\ \emph {et~al.}(2014)\citenamefont {Uola},
  \citenamefont {Moroder},\ and\ \citenamefont {Gühne}}]{UolaSteerJM1}%
  \BibitemOpen
  \bibfield  {author} {\bibinfo {author} {\bibfnamefont {R.}~\bibnamefont
  {Uola}}, \bibinfo {author} {\bibfnamefont {T.}~\bibnamefont {Moroder}}, \
  and\ \bibinfo {author} {\bibfnamefont {O.}~\bibnamefont {Gühne}},\ }\href
  {\doibase 10.1103/physrevlett.113.160403} {\bibfield  {journal} {\bibinfo
  {journal} {Physical Review Letters}\ }\textbf {\bibinfo {volume} {113}}
  (\bibinfo {year} {2014}),\ 10.1103/physrevlett.113.160403}\BibitemShut
  {NoStop}%
\bibitem [{\citenamefont {Uola}\ \emph {et~al.}(2015)\citenamefont {Uola},
  \citenamefont {Budroni}, \citenamefont {Gühne},\ and\ \citenamefont
  {Pellonpää}}]{UolaSteerJM2}%
  \BibitemOpen
  \bibfield  {author} {\bibinfo {author} {\bibfnamefont {R.}~\bibnamefont
  {Uola}}, \bibinfo {author} {\bibfnamefont {C.}~\bibnamefont {Budroni}},
  \bibinfo {author} {\bibfnamefont {O.}~\bibnamefont {Gühne}}, \ and\ \bibinfo
  {author} {\bibfnamefont {J.-P.}\ \bibnamefont {Pellonpää}},\ }\href
  {\doibase 10.1103/physrevlett.115.230402} {\bibfield  {journal} {\bibinfo
  {journal} {Physical Review Letters}\ }\textbf {\bibinfo {volume} {115}}
  (\bibinfo {year} {2015}),\ 10.1103/physrevlett.115.230402}\BibitemShut
  {NoStop}%
\bibitem [{\citenamefont {Kiukas}\ \emph {et~al.}(2017)\citenamefont {Kiukas},
  \citenamefont {Budroni}, \citenamefont {Uola},\ and\ \citenamefont
  {Pellonpää}}]{KiukasSteerJM}%
  \BibitemOpen
  \bibfield  {author} {\bibinfo {author} {\bibfnamefont {J.}~\bibnamefont
  {Kiukas}}, \bibinfo {author} {\bibfnamefont {C.}~\bibnamefont {Budroni}},
  \bibinfo {author} {\bibfnamefont {R.}~\bibnamefont {Uola}}, \ and\ \bibinfo
  {author} {\bibfnamefont {J.-P.}\ \bibnamefont {Pellonpää}},\ }\href
  {\doibase 10.1103/physreva.96.042331} {\bibfield  {journal} {\bibinfo
  {journal} {Physical Review A}\ }\textbf {\bibinfo {volume} {96}} (\bibinfo
  {year} {2017}),\ 10.1103/physreva.96.042331}\BibitemShut {NoStop}%
\bibitem [{\citenamefont {Jones}\ \emph {et~al.}(2022)\citenamefont {Jones},
  \citenamefont {Uola}, \citenamefont {Cope}, \citenamefont {Ioannou},
  \citenamefont {Designolle}, \citenamefont {Sekatski},\ and\ \citenamefont
  {Brunner}}]{jones2022}%
  \BibitemOpen
  \bibfield  {author} {\bibinfo {author} {\bibfnamefont {B.~D.}\ \bibnamefont
  {Jones}}, \bibinfo {author} {\bibfnamefont {R.}~\bibnamefont {Uola}},
  \bibinfo {author} {\bibfnamefont {T.}~\bibnamefont {Cope}}, \bibinfo {author}
  {\bibfnamefont {M.}~\bibnamefont {Ioannou}}, \bibinfo {author} {\bibfnamefont
  {S.}~\bibnamefont {Designolle}}, \bibinfo {author} {\bibfnamefont
  {P.}~\bibnamefont {Sekatski}}, \ and\ \bibinfo {author} {\bibfnamefont
  {N.}~\bibnamefont {Brunner}},\ }\href@noop {} {\bibfield  {journal} {\bibinfo
   {journal} {arXiv preprint arXiv:2207.04080}\ } (\bibinfo {year}
  {2022})}\BibitemShut {NoStop}%
\bibitem [{\citenamefont {Ioannou}\ \emph {et~al.}(2022)\citenamefont
  {Ioannou}, \citenamefont {Sekatski}, \citenamefont {Designolle},
  \citenamefont {Jones}, \citenamefont {Uola},\ and\ \citenamefont
  {Brunner}}]{ioannou2022}%
  \BibitemOpen
  \bibfield  {author} {\bibinfo {author} {\bibfnamefont {M.}~\bibnamefont
  {Ioannou}}, \bibinfo {author} {\bibfnamefont {P.}~\bibnamefont {Sekatski}},
  \bibinfo {author} {\bibfnamefont {S.}~\bibnamefont {Designolle}}, \bibinfo
  {author} {\bibfnamefont {B.~D.~M.}\ \bibnamefont {Jones}}, \bibinfo {author}
  {\bibfnamefont {R.}~\bibnamefont {Uola}}, \ and\ \bibinfo {author}
  {\bibfnamefont {N.}~\bibnamefont {Brunner}},\ }\href {\doibase
  10.1103/PhysRevLett.129.190401} {\bibfield  {journal} {\bibinfo  {journal}
  {Phys. Rev. Lett.}\ }\textbf {\bibinfo {volume} {129}},\ \bibinfo {pages}
  {190401} (\bibinfo {year} {2022})}\BibitemShut {NoStop}%
\bibitem [{\citenamefont {Terhal}\ and\ \citenamefont
  {Horodecki}(2000)}]{Terhal}%
  \BibitemOpen
  \bibfield  {author} {\bibinfo {author} {\bibfnamefont {B.~M.}\ \bibnamefont
  {Terhal}}\ and\ \bibinfo {author} {\bibfnamefont {P.}~\bibnamefont
  {Horodecki}},\ }\href {\doibase 10.1103/PhysRevA.61.040301} {\bibfield
  {journal} {\bibinfo  {journal} {Phys. Rev. A}\ }\textbf {\bibinfo {volume}
  {61}},\ \bibinfo {pages} {040301} (\bibinfo {year} {2000})}\BibitemShut
  {NoStop}%
\bibitem [{\citenamefont {Sanpera}\ \emph {et~al.}(2001)\citenamefont
  {Sanpera}, \citenamefont {Bru\ss{}},\ and\ \citenamefont
  {Lewenstein}}]{Sanpera}%
  \BibitemOpen
  \bibfield  {author} {\bibinfo {author} {\bibfnamefont {A.}~\bibnamefont
  {Sanpera}}, \bibinfo {author} {\bibfnamefont {D.}~\bibnamefont {Bru\ss{}}}, \
  and\ \bibinfo {author} {\bibfnamefont {M.}~\bibnamefont {Lewenstein}},\
  }\href {\doibase 10.1103/PhysRevA.63.050301} {\bibfield  {journal} {\bibinfo
  {journal} {Phys. Rev. A}\ }\textbf {\bibinfo {volume} {63}},\ \bibinfo
  {pages} {050301} (\bibinfo {year} {2001})}\BibitemShut {NoStop}%
\bibitem [{\citenamefont {Heinosaari}\ \emph {et~al.}(2015)\citenamefont
  {Heinosaari}, \citenamefont {Kiukas}, \citenamefont {Reitzner},\ and\
  \citenamefont {Schultz}}]{HeinosaariIBC}%
  \BibitemOpen
  \bibfield  {author} {\bibinfo {author} {\bibfnamefont {T.}~\bibnamefont
  {Heinosaari}}, \bibinfo {author} {\bibfnamefont {J.}~\bibnamefont {Kiukas}},
  \bibinfo {author} {\bibfnamefont {D.}~\bibnamefont {Reitzner}}, \ and\
  \bibinfo {author} {\bibfnamefont {J.}~\bibnamefont {Schultz}},\ }\href
  {\doibase 10.1088/1751-8113/48/43/435301} {\bibfield  {journal} {\bibinfo
  {journal} {Journal of Physics A: Mathematical and Theoretical}\ }\textbf
  {\bibinfo {volume} {48}},\ \bibinfo {pages} {435301} (\bibinfo {year}
  {2015})}\BibitemShut {NoStop}%
\bibitem [{\citenamefont {Heinosaari}\ \emph {et~al.}(2016)\citenamefont
  {Heinosaari}, \citenamefont {Miyadera},\ and\ \citenamefont
  {Ziman}}]{HeinosaariIncompatibilityReview}%
  \BibitemOpen
  \bibfield  {author} {\bibinfo {author} {\bibfnamefont {T.}~\bibnamefont
  {Heinosaari}}, \bibinfo {author} {\bibfnamefont {T.}~\bibnamefont
  {Miyadera}}, \ and\ \bibinfo {author} {\bibfnamefont {M.}~\bibnamefont
  {Ziman}},\ }\href {\doibase 10.1088/1751-8113/49/12/123001} {\bibfield
  {journal} {\bibinfo  {journal} {Journal of Physics A: Mathematical and
  Theoretical}\ }\textbf {\bibinfo {volume} {49}},\ \bibinfo {pages} {123001}
  (\bibinfo {year} {2016})}\BibitemShut {NoStop}%
\bibitem [{\citenamefont {G\"uhne}\ \emph {et~al.}(2023)\citenamefont
  {G\"uhne}, \citenamefont {Haapasalo}, \citenamefont {Kraft}, \citenamefont
  {Pellonp\"a\"a},\ and\ \citenamefont {Uola}}]{GuhneJMColloquium}%
  \BibitemOpen
  \bibfield  {author} {\bibinfo {author} {\bibfnamefont {O.}~\bibnamefont
  {G\"uhne}}, \bibinfo {author} {\bibfnamefont {E.}~\bibnamefont {Haapasalo}},
  \bibinfo {author} {\bibfnamefont {T.}~\bibnamefont {Kraft}}, \bibinfo
  {author} {\bibfnamefont {J.-P.}\ \bibnamefont {Pellonp\"a\"a}}, \ and\
  \bibinfo {author} {\bibfnamefont {R.}~\bibnamefont {Uola}},\ }\href {\doibase
  10.1103/RevModPhys.95.011003} {\bibfield  {journal} {\bibinfo  {journal}
  {Rev. Mod. Phys.}\ }\textbf {\bibinfo {volume} {95}},\ \bibinfo {pages}
  {011003} (\bibinfo {year} {2023})}\BibitemShut {NoStop}%
\bibitem [{\citenamefont {Hirsch}\ \emph {et~al.}(2013)\citenamefont {Hirsch},
  \citenamefont {Quintino}, \citenamefont {Bowles},\ and\ \citenamefont
  {Brunner}}]{hirsch2013}%
  \BibitemOpen
  \bibfield  {author} {\bibinfo {author} {\bibfnamefont {F.}~\bibnamefont
  {Hirsch}}, \bibinfo {author} {\bibfnamefont {M.~T.}\ \bibnamefont
  {Quintino}}, \bibinfo {author} {\bibfnamefont {J.}~\bibnamefont {Bowles}}, \
  and\ \bibinfo {author} {\bibfnamefont {N.}~\bibnamefont {Brunner}},\
  }\href@noop {} {\bibfield  {journal} {\bibinfo  {journal} {Physical review
  letters}\ }\textbf {\bibinfo {volume} {111}},\ \bibinfo {pages} {160402}
  (\bibinfo {year} {2013})}\BibitemShut {NoStop}%
\bibitem [{\citenamefont {Zeng}\ \emph {et~al.}(2018)\citenamefont {Zeng},
  \citenamefont {Wang}, \citenamefont {Li},\ and\ \citenamefont
  {Zhang}}]{Zeng2018}%
  \BibitemOpen
  \bibfield  {author} {\bibinfo {author} {\bibfnamefont {Q.}~\bibnamefont
  {Zeng}}, \bibinfo {author} {\bibfnamefont {B.}~\bibnamefont {Wang}}, \bibinfo
  {author} {\bibfnamefont {P.}~\bibnamefont {Li}}, \ and\ \bibinfo {author}
  {\bibfnamefont {X.}~\bibnamefont {Zhang}},\ }\href {\doibase
  10.1103/PhysRevLett.120.030401} {\bibfield  {journal} {\bibinfo  {journal}
  {Phys. Rev. Lett.}\ }\textbf {\bibinfo {volume} {120}},\ \bibinfo {pages}
  {030401} (\bibinfo {year} {2018})}\BibitemShut {NoStop}%
\bibitem [{\citenamefont {Srivastav}\ \emph {et~al.}(2022)\citenamefont
  {Srivastav}, \citenamefont {Valencia}, \citenamefont {McCutcheon},
  \citenamefont {Leedumrongwatthanakun}, \citenamefont {Designolle},
  \citenamefont {Uola}, \citenamefont {Brunner},\ and\ \citenamefont
  {Malik}}]{Srivastav2022}%
  \BibitemOpen
  \bibfield  {author} {\bibinfo {author} {\bibfnamefont {V.}~\bibnamefont
  {Srivastav}}, \bibinfo {author} {\bibfnamefont {N.~H.}\ \bibnamefont
  {Valencia}}, \bibinfo {author} {\bibfnamefont {W.}~\bibnamefont
  {McCutcheon}}, \bibinfo {author} {\bibfnamefont {S.}~\bibnamefont
  {Leedumrongwatthanakun}}, \bibinfo {author} {\bibfnamefont {S.}~\bibnamefont
  {Designolle}}, \bibinfo {author} {\bibfnamefont {R.}~\bibnamefont {Uola}},
  \bibinfo {author} {\bibfnamefont {N.}~\bibnamefont {Brunner}}, \ and\
  \bibinfo {author} {\bibfnamefont {M.}~\bibnamefont {Malik}},\ }\href
  {\doibase 10.1103/PhysRevX.12.041023} {\bibfield  {journal} {\bibinfo
  {journal} {Phys. Rev. X}\ }\textbf {\bibinfo {volume} {12}},\ \bibinfo
  {pages} {041023} (\bibinfo {year} {2022})}\BibitemShut {NoStop}%
\bibitem [{\citenamefont {Hayashi}(1998)}]{hayashi1998}%
  \BibitemOpen
  \bibfield  {author} {\bibinfo {author} {\bibfnamefont {M.}~\bibnamefont
  {Hayashi}},\ }\href@noop {} {\bibfield  {journal} {\bibinfo  {journal}
  {Journal of Physics A: Mathematical and General}\ }\textbf {\bibinfo {volume}
  {31}},\ \bibinfo {pages} {4633} (\bibinfo {year} {1998})}\BibitemShut
  {NoStop}%
\end{thebibliography}%

\begin{widetext}
\appendix

\section{Proof of Lemma 1}

\label{app: lemma 1}

\noindent \textbf{Lemma 1}:  \textit{Consider an assemblage $\{\sigma_{a|x}\}_{a,x}$ and a loss channel $\cL_\eta$ for $\eta>0$. The assemblage  $\left\{\sigma'_{a|x}=  \cL_\eta[\sigma_{a|x}] \right\}_{a,x} $ is $n$-steerable if and only if $\{\sigma_{a|x}\}_{a,x}$ is.}\\

First let us prove that if $\{\sigma_{a|x}\}_{a,x}$ is not $n$-steerable than $\{\sigma_{a|x}'\}_{a,x}$ is neither.  Here, $\{\sigma_{a|x}\}_{a,x}$ is not $n$-steerable means that there exist a rank-$n$ hidden state (R$n$HS) model which allows to simulate the assemblage. Such a model consists of a probability density $p(\lambda)$ over states $\psi_{AB}(\lambda)$ that all have Schmidt rank at most $n$, and a collection of "response POVMs" $A_{a|x,\lambda}$ performed by Alice, such that 
\be
\sigma_{a|x} = \int \dd\lambda p(\lambda) \tr_A \psi_{AB}(\lambda) A_{a|x ,\lambda}.
\ee
It is straightforward to define R$n$HS model for the assemblage $\{\sigma_{a|x}'\}_{a,x}$ by simply applying the loss channel to each state $\psi'_{AB}(\lambda) = \t{id}_A\otimes \cL_\eta [\psi_{AB}(\lambda)]$. Note that the loss channel (as any local map) cannot increase the Schmidt rank of a state and
\be\begin{split}
 \int \dd\lambda p(\lambda) \tr_A (\t{id}_A\otimes \cL_\eta [\psi_{AB}(\lambda)]) A_{a|x ,\lambda}= \cL_\eta \left[ \int \dd\lambda p(\lambda) \tr_A \psi_{AB}(\lambda) A_{a|x ,\lambda} \right]
= \cL[\sigma_{a|x}] =\sigma_{a|x}'.    
\end{split}
\ee

Next, we prove the other direction -- if $\{\sigma_{a|x}'\}_{a,x}$ is not $n$-steerable then $\{\sigma_{a|x}\}_{a,x}$ is neither. Now we start with a R$n$HS model for the assemblage model $p'(\lambda)$, $\psi_{AB}'(\lambda)$ and $A_{a|x,\lambda}$ for the assemblage 
\be
\sigma_{a|x}' = \int \dd \lambda p'(\lambda) \tr_A \psi'_{AB}(\lambda) A_{a|x,\lambda}.
\ee
The states $\sigma_{a|x}'$ and $\phi_{AB}'(\lambda)$ are here defined on a $(d+1)$-dimensional system involving an additional level $\ket{\nc}$ introduced by the loss channel. Let us define the restriction of these states on the subspace orthogonal to $\ket{\psi}$ projected by $\Pi_{[1,d]}$. For the hidden state we obtain
\be
\psi_{AB}(\lambda)\equiv \frac{(\t{id}_A\otimes \Pi_{[1,d]}) \psi_{AB}'(\lambda) (\t{id}_A\otimes \Pi_{[1,d]})}{q_{[1,d]}(\lambda)} \qquad \text{with} \qquad q_{[1,d]}(\lambda)= \tr (\t{id}_A\otimes \Pi_{[1,d]}) \psi_{AB}'(\lambda) 
\ee
that also have Schmidt rank at most $n$, and define new probability density
\be
p(\lambda) \equiv \frac{p'(\lambda) q_{[1,d]}(\lambda) }{\int \dd \lambda p'(\lambda) q_{[1,d]}(\lambda)}.
\ee
Here the denominator is 
\be
\int \dd \lambda p'(\lambda) q_{[1,d]}(\lambda) = \tr (\t{id}_A\otimes \Pi_{[1,d]}) \int \dd \lambda p'(\lambda) \psi'_{AB}(\lambda) = \tr \Pi_{[1,d]}\sum_{a}\sigma_{a|x}' = \eta
\ee
Now compute which assemblage the novel R$n$HS model $(p(\lambda),\psi_{AB}(\lambda), A_{a|x,\lambda})$ simulates
\be\begin{split}
\tilde \sigma_{a|x}&=\int \dd \lambda p(\lambda) tr_A \psi_{AB}(\lambda) (\t{id}_A\otimes \Pi_{[1,d]}) A_{a|x,\lambda} \\
& = \frac{\int \dd \lambda p'(\lambda) tr_A  (\t{id}_A\otimes \Pi_{[1,d]}) \psi'_{AB}(\lambda) (\t{id}_A\otimes \Pi_{[1,d]}) A_{a|x,\lambda}}{\int \dd \lambda p'(\lambda) q_{[1,d]}(\lambda)}  \\
& = \frac{ \Pi_{[1,d]}\left(\int \dd \lambda p'(\lambda) tr_A   \psi'_{AB}(\lambda)  A_{a|x,\lambda}\right) \Pi_{[1,d]}}{\eta } \\
& = \frac{1}{\eta} \Pi_{[1,d]} \sigma_{a|x}'\ \Pi_{[1,d]}\\
& = \sigma_{a|x}
\end{split}
\ee
To show the last equality consider the action of  loss channel on $\sigma_{a|x}$
\be
\sigma_{a|x}'= \cL_\eta(\sigma_{a|x}) = \eta \,  \sigma_{a|x} + (1-\eta) \ketbra{\nc} \implies \sigma_{a|x} = \frac{1}{\eta} \Pi_{[1,d]} \sigma_{a|x}' \Pi_{[1,d]}.
\ee
This concludes the proof. $\square$


\section{Computing the integrals in the proof of the Result 2}
\label{app: integrals}
Here we compute that integrals 
\be\begin{split}
\bar A_d(t) &=  d \int \dd \bm z\,  p_t^{(a)}(a|\bm z)\, \left|\braket{\varphi_a}{\bm z}\right|^2 \\
\bar T_d(t)  &= d \int \dd \bm z\,  p_t^{(a)}(a|\bm z),
\end{split}
\ee
where $\dd \bm z$ is the unitary-invariant measure over quantum state $\ket{\bm z}$ and  $p_t^{(a)}(a|\bm z) = \begin{cases} 1 & |\braket{\varphi_a}{\bm z}|^2\geq t\\ 
0 & \text{otherwise}
\end{cases}$.
It will be convening to work in the basis where  $\ket{\varphi_a}=\ket{0}$, since the measure is invariant we directly get
\be\begin{split}
\bar A_d(t) &=  d \int \dd \bm z\,  \Xi[\bm z] \, \left|\braket{0}{\bm z}\right|^2 \\
\bar T_d(t)  &= d \int \dd \bm z\, \Xi[\bm z]  ,
\end{split}
\ee
and $\Xi_t(\bm z) = \begin{cases} 1 & |\braket{0}{\bm z}|^2\geq t\\ 
0 & \text{otherwise}
\end{cases}$. To compute these integrals we parametrize the quantum states $\ket{\bm z}=\sum_{k=0}^{d-1} z_k \ket{k}$ with
\be
\bm z =\left(\begin{array}{c}
z_0\\
z_1\\
\vdots\\
z_{d-2}\\
z_{d-1}
\end{array}\right) 
= \left( \begin{array}{c} \cos(\theta_1) \\
e^{\ii \phi_2} \sin(\theta_1) \cos(\theta_2)\\
\vdots\\
e^{\ii \phi_{d-1}} \sin(\theta_1)\dots \sin(\theta_{d-2}) \cos(\theta_{d-1}) \\
e^{\ii \phi_{d}} \sin(\theta_1)\dots \sin(\theta_{d-2}) \sin(\theta_{d-1})
\end{array}\right)
\label{eq: complex z}
\ee
for $\phi_i \in [0,2\pi] $ and $\theta_i \in [0,\frac{\pi}{2}]$, for which the invariant measure takes the form~\cite{hayashi1998} 
\be\label{eq: measure states}
\dd \bm z = \frac{(d-1)!}{ \pi^{d-1}} \Pi_{i=1}^{d-1}\sin^{2d-2i-1}(\theta_i) \cos(\theta_i)\dd \theta_i\,  \Pi_{j=2}^d\dd\phi_j.
\ee
For such parametrization we obtain $|\braket{0}{\bm z}|^2 = |z_0|^2= \cos^2(\theta_1)$ and 
\be
\Xi_t(\bm z) \equiv \Xi_t(\theta_1) = \begin{cases} 1 &\cos^2(\theta_1) \geq t\\ 
0 & \text{otherwise}.
\end{cases}
\ee 
The integrand only depends on the angle $\theta_1$, which allow us to write
\be\begin{split}
\bar T_d(t) &= \frac{d}{N} \int_0^{\pi/2}   \sin^{2d-3}(\theta_1) \cos(\theta_1)\dd\theta_1 \Xi_t(\theta_1) \\
\bar A_d(t) &= \frac{d}{N} \int_0^{\pi/2}  \sin^{2d-3}(\theta_1) \cos(\theta_1) \dd\theta_1 \Xi_t(\theta_1) \cos^2(\theta_1).    
\end{split}
\ee
with the normalization constant $N = \int_0^{\pi/2} \sin^{2d-3}(\theta_1) \cos(\theta_1)\dd\theta_1$. Perform the variable change $x=\sin^2(\theta_1)$ with ${\dd \theta_1}  = \frac{\dd x}{2\sin(\theta_1)\cos(\theta_1)}$, we obtain the expressions used in the main text
\be\begin{split}
N &= \frac{1}{2}\int_0^{1} x^{d-2}\dd x = \frac{1}{2(d-1)}\\
\bar T_d(t) &= \frac{d}{2N} \int_0^{1-t} x^{d-2}\dd x = d(1-t)^{d-1}\\
\bar A_d(t) &= \frac{d}{2N} \int_0^{1-t} x^{d-2}(1-x)\dd x = (1-t)^{d-1} ((d-1) t+1).   
\end{split}
\ee


\section{Proof of Lemma 2}

\label{app: lemma 2}

 \noindent \textbf{Lemma 2} {\it The channel $\cE=\cL_\eta\circ \cW_p$ is incompatibility breaking if $\eta \leq (1-p)^{d-1}$}.\\

\textit{Proof}: By Result 3 we know that all POVMs $\{\bar M_{a|U}^{(\eta,p)}\}_U$ in dimension $d$ subject to white noise and loss are jointly measurable if $\eta \leq (1-p)^{d-1}$. Lemma 2 is thus implied by the following result.\\

\noindent \textbf{Lemma 2.1} {\it The channel $\cE=\cL_\eta\circ \cW_p$ is incompatibility breaking if and only if the set of all POVMs $\{\bar M_{a|x}^{(\eta,p)}\}$ subject to loss and white noise is jointly measurable.}\\

\textit{Proof:} This claim might seem a tautology at first glance, but at closer inspection one realises that the channel $\cE$ outputs $(d+1)$-dimensional, while $\bar M_{a|x}$ describe measurements of a $d$-dimensional system. So the predicate does not refer to the same sets of POVMs.

More precisely, let the set $\{M_{a|x}'\}$ denote all POVMs acting in dimension $d+1$. Then $\cE$ is incompatibility breaking iff the measurement assemblage
\be
\{ \hat{M} _{a|x}^{(\eta,p)} \equiv \cE^*[M'_{a|x}]\} 
\ee
is jointly measurable. While the set $\{\bar M_{a|x}^{(\eta,p)}\}$ contains the POVMs
\be
\bar M_a^{(\eta,p)} = 
\begin{cases}
 \eta p \, M_{a} + \eta (1-p) (\tr M_a) \frac{\id_d}{d} & a=1,\dots,m  \\ 
 (1-\eta) \id_d & a= \nc
\end{cases}
\ee
where $M_{a|x}$ runs through all POVMs in dimension $d$. We want to show that the assemblage $\{\hat M_{a|x}^{(\eta,p)}\}$ is compatible if and only if $\{\bar M_{a|x}^{(\eta,p)}\}$ is, where both sets contain measurements acting on a $d$-dimensional quantum system.

To start let us have a closer look on the assemblage $\{\hat M_{a|x}\}$.  The loss channel $\cL_\eta$ has $d+1$ Kraus operators 
\be
\begin{split}
K_\checkmark &= \sqrt{\eta} \underbrace{\sum_{k=1}^d \ketbra{k}}_{\equiv \Pi_{\checkmark}} \\
K_k &= \sqrt{1-\eta} \ketbra{\nc}{k}    \qquad k =0,\dots,d-1 
\end{split}
\ee
Thus, for any measurement $M_a'$ on the Hilbert space $\cH_{d+1}=\text{span}
\{\ket{0},\dots,\ket{d-1},\ket{\nc}\}$ we have $\cE^*[M_a']= \cW_p^* \circ \cL_\eta^* [M_a']$ and 
\be\begin{split}
\cL_\eta^*[M_a'] &= K_\checkmark^\dag M_a' K_\checkmark + \sum_{k=0}^{d-1} K_k^\dag M_a' K_k\\
    & = \eta \, \Pi_\checkmark M_a' \Pi_\checkmark + (1-\eta) \id_d \underbrace{\bra{\nc} M_a' \ket{\nc}}_{\equiv q(a)},
\end{split}
\ee
where $q(a)$ is a probability distribution since $q(a) = \tr \ketbra{\nc } M_a'\geq 0$ and $\sum_a q(a) = \tr \ketbra{\nc } \sum_a M_a' = 1$. In addition, the operators $M_a \equiv \Pi_\checkmark M_a' \Pi_\checkmark$ define a valid POVM on the $d$-dimensional subspace $\cH_{d}=\text{span}
\{\ket{0},\dots,\ket{d-1}\}$ prior to losses. We thus have
\be
\begin{split}
    \cL_\eta^*[M_a'] = \eta M_a + (1-\eta) q (a)  \id_d.
\end{split}
\ee
Finally adding the noise channel we get 
\be\label{app: last smth}\begin{split}
  \hat M_a^{(\eta,p)}
  &= \cW_p^*\circ \cL_\eta^*[M_a'] 
  \\
  &= \eta \cW_p^* [M_a] + (1-\eta)q(a) \cW_p^*[\id_d] \\    
  & = \eta M_a^{(1,p)} + (1-\eta)q(a) \id_d \\
  & = \eta p M_a + \eta(1-p) (\tr M_a) \frac{\id_d}{d} + (1-\eta)q(a) \id_d \\
  & = \bar M_a^{(\eta,p)} +q(a) \bar M_\nc^{(\eta,p)}.
\end{split}
\ee
We see that if the POVM $\bar M_a^{(\eta,p)}$ can be simulated, one can also use it to simulate $\hat M_a^{(\eta,p)}$ by simply relabeling the output $\nc$ to $a$ with probability $q(a)$. Hence if the set of all POVMs  $\{\bar M_{a|x}^{(\eta,p)}\}$ is compatible this is also the case of $\{\hat M_{a|x}^{(\eta,p)}\}$.

To show the converse consider all POVMs on the $(d+1)$-dimensional space with the elements of the form $\{M_1,\dots,M_a,\dots, M_\nc=\ketbra{\nc}\}$ with $\bra{\nc} M_a \ket{\nc}=0$, and $\{M_1,\dots,M_a,\dots \}$ (without the $M_\nc$ element) running through all POVMs on the $d$-dimensional subspace $\cH_{d}=\text{span}
\{\ket{0},\dots,\ket{d-1}\}$. For a POVM of such form we have $q(a)=0$, $q(\nc)=1$ and by virtue of Eq.~\label{app: last smth}
\be
\begin{split}
    \hat M_a^{(\eta,p)} &=  \bar M_a^{(\eta,p)} \\
    \hat M_\nc^{(\eta,p)} &=  (1-\eta) \bar M_\nc^{(\eta,p)}.
\end{split}
\ee
This if any measuremnt $\hat M_a^{(\eta,p)}$ can be simulated any measurement $\bar M_a^{(\eta,p)}$ can be as well. $\square$

\end{widetext}
\end{document}